\documentclass[twoside]{article}

\usepackage{hmatrix_format}
\usepackage{verbatim}
\usepackage{graphicx}
\usepackage{xcolor}

\usepackage[numbers]{natbib}
\usepackage{microtype}

\usepackage[hide=false,setmargin=true,marginparwidth=0.5in]{marginalia}

\usepackage{multicol}
\usepackage{}

\usepackage{graphicx} 
\usepackage{subfigure}

\usepackage{url}
\usepackage{bm}
\usepackage{framed}

\usepackage{amsmath}
\usepackage{amsfonts}
\DeclareMathAlphabet{\mathpzc}{OT1}{pzc}{m}{it}

\usepackage{stackrel}

\usepackage[shortlabels]{enumitem}

\usepackage{xcolor}
\usepackage{tikz}

\usepackage{rotating}

\usepackage{amsthm}
\usepackage{mathtools}

\usepackage{algorithm}
\usepackage{algpseudocode}

\usepackage[overlay]{textpos}

\theoremstyle{theorem}

\theoremstyle{lemma}

\theoremstyle{definition}

\theoremstyle{conjecture}

\usetikzlibrary{calc,trees,positioning,arrows,chains,shapes.geometric,%
    decorations.pathreplacing,decorations.pathmorphing,shapes,%
    matrix,shapes.symbols}

\tikzset{
>=stealth',
  punktchain/.style={
    rectangle, 
    rounded corners, 
    draw=black, very thick,
    text width=10em, 
    minimum height=3em, 
    text centered, 
    on chain},
  line/.style={draw, thick, <-},
  element/.style={
    tape,
    top color=white,
    bottom color=blue!50!black!60!,
    minimum width=8em,
    draw=blue!40!black!90, very thick,
    text width=10em, 
    minimum height=3.5em, 
    text centered, 
    on chain},
  every join/.style={->, thick,shorten >=1pt},
  decoration={brace},
  tuborg/.style={decorate},
  tubnode/.style={midway, right=2pt},
}

\usepackage{hyperref}

\usepackage{dblfloatfix}
\usepackage{float}
\usepackage{inconsolata}

\usepackage{hyperref}
\hypersetup{
  colorlinks=true,
  citecolor=blue,
  urlcolor=magenta,
}
\begin{document}

\twocolumn[
  \ourtitle{Coarse-Grained Nonlinear System Identification}
  \ourauthor{Span Spanbauer \And Ian Hunter}
  \ouraddress{MIT \And MIT}

]


\begin{abstract}
We introduce \textit{Coarse-Grained Nonlinear Dynamics}, an efficient and universal parameterization of nonlinear system dynamics based on the Volterra series expansion. These models require a number of parameters only quasilinear in the system's memory regardless of the order at which the Volterra expansion is truncated; this is a superpolynomial reduction in the number of parameters as the order becomes large. This efficient parameterization is achieved by coarse-graining parts of the system dynamics that depend on the product of temporally distant input samples; this is conceptually similar to the coarse-graining that the fast multipole method uses to achieve $\mathcal{O}(n)$ simulation of n-body dynamics. Our efficient parameterization of nonlinear dynamics can be used for regularization, leading to \textit{Coarse-Grained Nonlinear System Identification}, a technique which requires very little experimental data to identify accurate nonlinear dynamic models. We demonstrate the properties of this approach on a simple synthetic problem. We also demonstrate this approach experimentally, showing that it identifies an accurate model of the nonlinear voltage to luminosity dynamics of a tungsten filament with less than a second of experimental data.

\end{abstract}



\section{Introduction}

The Volterra series expansion is a foundational method for modeling nonlinear dynamics. Since its introduction in 1887~\citep{volterra1887}, it, along with its orthonormalized counterpart the Wiener series expansion~\citep{wiener1966nonlinear}, have seen widespread application. It has been applied extensively in biology~\citep{korenberg1996identification,korenberg1990identification}, for example to model the human pupillary light reflex~\citep{sandberg1968wiener}, the retina's receptive-field response~\citep{marmarelis1973nonlinear}, spike-train responses of neurons~\citep{joeken1997modeling}, and human skin mechanics~\citep{chen2013nonlinear}. It has been applied in medical and chemical instrumentation~\citep{litt1997nonlinear} such as CT and NMR~\citep{asfour2000nonlinear}, in electronics and electromechanical systems~\citep{zhu2004behavioral,narayanan1970application,kaizer1987modeling}, in optical systems such as LED and laser physics~\citep{biswas1991volterra,kamalakis2011empirical}, and in energy storage and transfer~\citep{gruber2012nonlinear,hunter2016apparatus,lafontaine2018nonlinear}.

Despite the success of the Volterra series expansion, the majority of these applications truncate the series at the first nonlinear term, that is, the second order Volterra kernel. This is primarily because of statistical difficulties in accurately estimating higher-order Volterra kernels. These difficulties arise due to the rapid exponential growth of the number of parameters in a Volterra series as the order of the series $d$ and the system memory $n$ increase; the number of parameters in a standard Volterra series scales as $\mathcal{O}(n^d)$ which lies in the multiparameter complexity class \textbf{XP}~\citep{downey2013fundamentals}. 

We achieve a superpolynomial reduction in this scaling to the strictly smaller multiparameter complexity class \textbf{FPT}~\citep{downey2013fundamentals} by employing coarse-graining inspired by the fast multipole method~\citep{rokhlin1985}, famous for its ability to simulate n-body dynamics in only $\mathcal{O}(n)$ time.

\noindent {\bf Contributions.} This paper presents a method of dramatically reducing the amount of data required to accurately estimate Volterra kernels without sacrificing useful representational power. Specifically, it presents the following contributions:
\vspace*{-5pt}
\begin{enumerate}
\item This paper introduces \textit{Coarse-Grained Nonlinear Dynamics}, a Volterra series-based method of universally approximating nonlinear dynamics achieving a superpolynomial reduction in the number of model parameters from $\mathcal{O}(n^d)$ to $\mathcal{O}(2^d \: d \: n \: \text{log}_2 \: n)$, which is a strict reduction from the complexity class \textbf{XP} to the complexity class \textbf{FPT}.
\item This paper introduces \textit{Coarse-Grained Nonlinear System Identification}, which uses this reduction in the number of model parameters as a statistical regularizer for extremely fast estimation of Volterra kernels.
\item This paper provides a synthetic example highlighting the statistical regularization properties of model parameter reduction in a simple setting.
\item This paper provides an experimental demonstration of \textit{Coarse-Grained Nonlinear System Identification}, applying it to learn the nonlinear voltage to luminosity dynamics of a tungsten filament. Our method achieves 99.4\% variance accounted for using only one second of training data, compared to the 97.5\% variance accounted for attained by a traditionally regularized Volterra model.
\end{enumerate}

\section{Background}
\label{sec:background}
\vspace{-5pt}

This section provides a brief introduction to the Volterra series expansion as well as to hierarchical matrices~\citep{hackbusch1999} and their generalization to higher rank tensors, which we use to implement temporal coarse-graining in our method.

\subsection{Volterra series expansions}

The Volterra series expansion is a universal approximator of nonlinear dynamics analogous to a multivariable Taylor series in a continuous range of variables, specifically the set of historical input values.

The coefficients of this series expansion for the terms of a given order $d$ are collected into real-valued arity-$d$ functions called Volterra kernels $h_d$, leading to the definition of the Volterra series expansion, shown here to second order, where $T$ is the memory of the system:
\begin{align*}
y(t) &= h_0 \\
&+ \int_{-T}^0 h_1(\tau_1)x(t-\tau_1) d\tau_1\\
&+ \int_{-T}^0 \int_{-T}^0 h_2(\tau_1,\tau_2)x(t-\tau_1)x(t-\tau_2) d\tau_1 d\tau_2\\
&+ \ldots
\end{align*}

The Volterra series expansion is a direct generalization of linear system dynamics. The first order term in the Volterra series is the convolution which describes linear dynamics in the time domain; the first order Volterra kernel $h_1$ is identified with the impulse response.

In practice the system memory $[-T,0]$ is discretized into $n$ samples; the arity-$d$ functions become rank-$d$ tensors of size $n$ in each dimension, hence the aforementioned $\mathcal{O}(n^d)$ scaling in the number of parameters.

\textit{Coarse-Grained Nonlinear Dynamics} achieves a superpolynomial reduction in the scaling with $n$ as $d$ becomes large; it achieves this by representing Volterra kernels with hierarchical tensors, implementing a form of temporal coarse-graining. We now turn to an overview of hierarchical matrices and their generalization to higher-rank tensors.

\begin{figure}[!t]
\centering
\begin{picture}(240,153)
\put(0,0){\includegraphics[width=82mm]{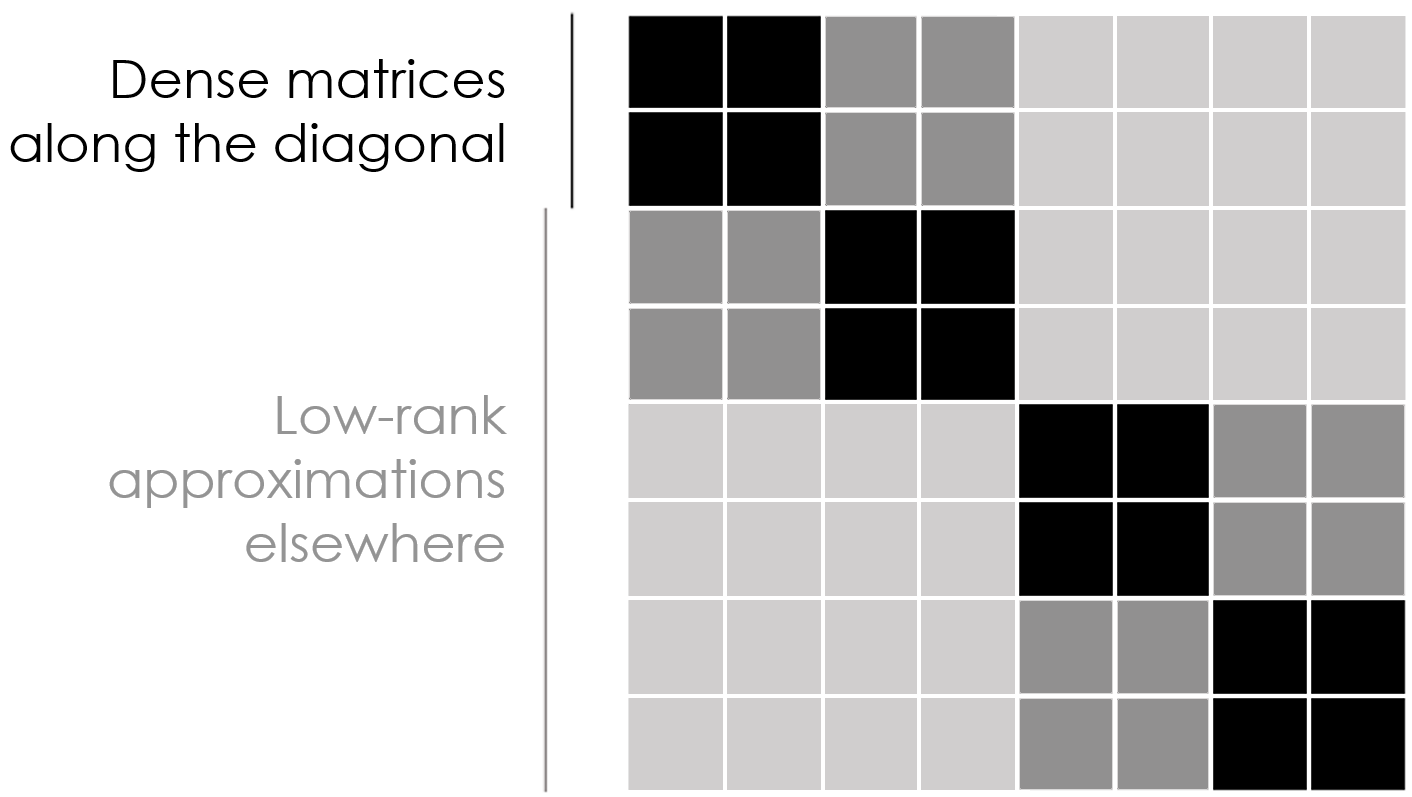}}
\put(32,144){\textbf{Structure of a hierarchical matrix}}

\end{picture}
\caption{In this diagram of the structure of a hierarchical matrix, entries along the diagonal are represented in a dense format, while off-diagonal entries are represented using a low-rank approximation. This format requires only $\mathcal{O}(n \: \text{log}_2 \: n)$ storage and concentrates representational power near the diagonal which, for us, implements a form of temporal coarse-graining.}
\label{fig:h-diagram}
\end{figure}

\subsection{Hierarchical matrices and tensors}

Hierarchical matrices, introduced in 1999~\citep{hackbusch1999}, implement a form of coarse graining tightly related to the panel clustering algorithm~\citep{hackbusch1990panel} which is widely used to accelerate boundary element method simulations, and the fast multipole method~\citep{rokhlin1985} which is well known for its ability to simulate the n-body problem in $\mathcal{O}(n)$ time.

Hierarchical matrices can be specified with only $\mathcal{O}(n \: \text{log}_2 \: n)$ parameters; we leverage this reduction in number of parameters for statistical regularization, structurally imposing a statistical prior that the parts of the system dynamics that depend on the product of temporally distant input samples can be coarse-grained using low-rank approximations.

The structure of a hierarchical matrix, illustrated in Fig.~\ref{fig:h-diagram}, can be described recursively. A $2^p \times 2^p$ hierarchical matrix is composed of four equally sized block submatrices; the off-diagonal quadrants are each specified by a $k$-low rank approximation, that is, the sum of $k$ two vector outer products, where $k$ can be adjusted to tune the representational power available in the off-diagonal blocks. The on-diagonal quadrants are each specified recursively by a $2^{p-1} \times 2^{p-1}$ hierarchical matrix. Observe that the number of parameters $N(p)$ required to specify a $2^p \times 2^p$ hierarchical matrix is bounded above by the following recurrence relation:
\begin{align*}
N(p) \: &= \: 4 \: k \: 2^{p-1} + 2N(p-1)\\
N(0) \: &= \: 1
\end{align*}
which has the solution
\begin{align*}
N(p)&=2^p \: (2 \: k \: p\: + \: 1)\\
&=n(2 \: k \: \text{log}_2 \: n \: + \: 1) \: \text{for }n=2^p \text{,}
\end{align*}
hence the aforementioned $\mathcal{O}(n \: \text{log}_2 \: n)$ scaling.

Hierarchical matrices can be generalized to $d$-rank tensors following a similar recursive definition. A $d$-rank hierarchical tensor of size $(2^p)^d$ is composed of $2^d$ equally sized subtensors; the $2^d-2$ off-diagonal subtensors are each specified by a $k$-low rank approximation, that is, the sum of $k$ outer products of $d$ vectors each. The two on-diagonal subtensors are each specified recursively by a size $(2^{p-1})^d$ $d$-rank hierarchical tensor. The number of parameters $N_d(p)$ required to specify a $d$-rank hierarchical tensor of size $(2^p)^d$ is bounded above by the following recurrence relation:
\begin{align*}
N_d(p) \: &= \: k \: d \: (2^d - 2) 2^p + 2N_d(p-1)\\
N_d(0) \: &= \: 1
\end{align*}
which has the solution
\begin{align*}
N(p)=&2^p (2^{d-1} d k p - d k p                 + 1)\\
    =&n   (2^{d-1} d k \: \text{log}_2 n - d k \: \text{log}_2 \: n + 1) \: \text{for }n=2^p \text{,}
\end{align*}
which scales as $\mathcal{O}(2^d \: d \: n \: \text{log}_2 \: n)$. This is a superpolynomial reduction of the scaling in $n$ as $d$ becomes large compared to that of a dense tensor $\mathcal{O}(n^d)$. In fact, the scaling of the number of parameters in a hierarchical tensor lies in a strictly smaller multiparameter complexity class \textbf{FPT} than that of a dense tensor, which lies in \textbf{XP}.

In Section~\ref{sec:synthetic} we use a simple synthetic example to highlight properties of the statistical regularization which arises from a reduction in the number of model parameters. Then in Section~\ref{sec:physical} we demonstrate \textit{Coarse-Grained Nonlinear System Identification} experimentally, showing how this regularization enables us to accurately learn the nonlinear voltage to luminosity dynamics of a tungsten filament with only one second of training data.

\section{Synthetic experiment}
\label{sec:synthetic}
In this section we demonstrate properties of statistical regularization arising from model parameter reduction. We consider the task of learning a discretized integral operator from a small number of noisy random samples of the transform.

\begin{figure}[!t]

\centering

\begin{picture}(174,174)
\put(0,-16){\includegraphics[width=60mm]{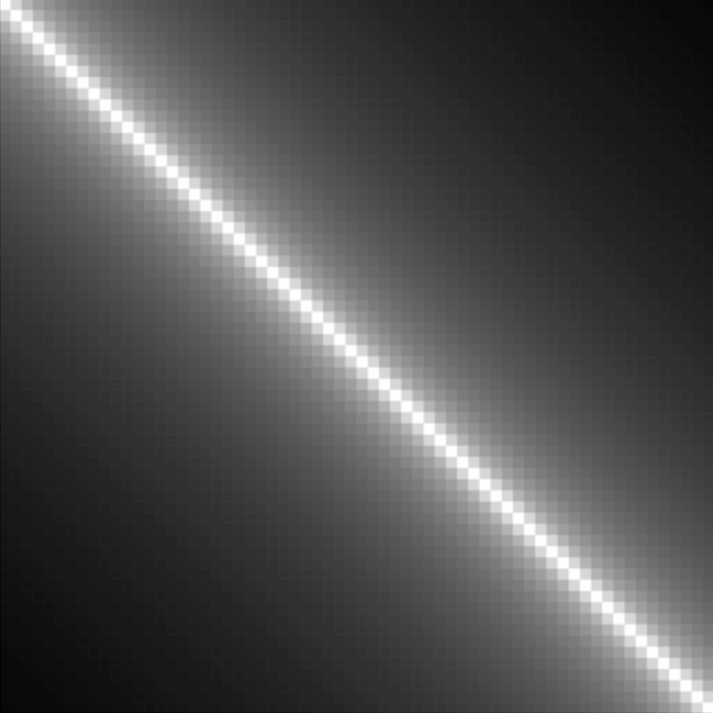}}
\put(12,164){\textbf{Discretized integral operator}}
\end{picture}

\vspace{5mm}

\caption{Matrix entries of the integral operator $\tilde f(t) = \int_0^1 \text{log}\:|t-s|\:f(s)\:ds$ discretized using 64 piecewise constant basis functions. We estimate the parameters of this operator from random noisy examples of this transform using various model classes and observe statistical regularization when using the model classes with the fewest parameters.}
\label{fig:intop}
\end{figure}

Specifically we attempt to learn a discretized form of the following operator:
$$
\tilde f(t) = \int_0^1 \text{log}\:|t-s|\:f(s)\:ds
$$
We discretize f(s) on [0,1] using piecewise constant basis functions
$$
f(s) = \sum_{j=1}^N f_j v_j(s)
$$
where $v_j$ is 1 on $\Gamma_j=[\frac{f-1}{N},\frac{j}{N}]$ and 0 elsewhere. Thus
$$
\tilde f(t) = \sum_{j=1}^N f_j \int_{\Gamma_j} \text{log}\:|t-s|\:f(s)\:ds .
$$
This integral can be evaluated analytically:
\begin{align*}
&\int_a^b \text{log}\:|t-s|\:f(s)\:ds\\
&= -(b-a)-\frac{1}{2}[(t-b)\text{log}(t-b)^2-(t-a)\text{log}(t-a)^2]
\end{align*}
which allows us to directly evaluate the entries in the matrix $A$ implementing the discretized form of the integral operator, as shown in Fig.~\ref{fig:intop}.
$$
\tilde f_j = Af_j
$$

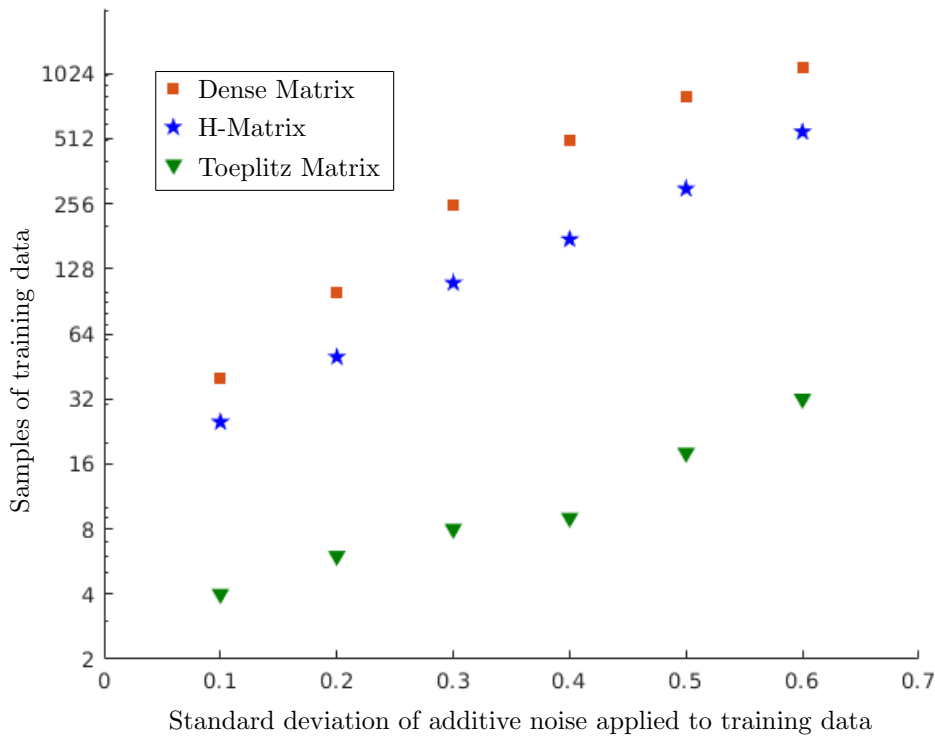
\begin{figure*}[!t]
\label{fig:bayesian}
\centering

\begin{picture}(370,280)
\put(0,0){\includegraphics[width=120mm]{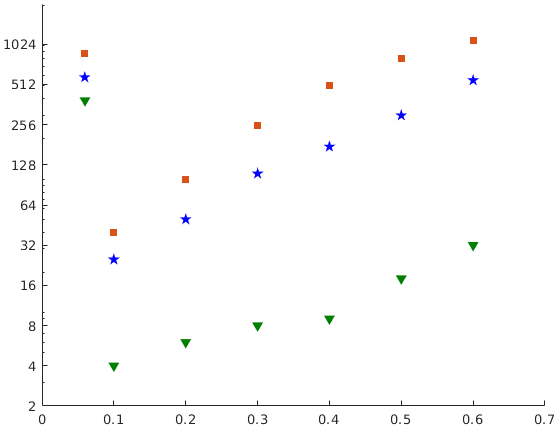}}
\put(-25,270){\textbf{Synthetic training data required to achieve a given accuracy in held-out data}}
\put(61,227){Dense Matrix}
\put(61,212.5){H-Matrix}
\put(61,197.3){Toeplitz Matrix}
\put(45,237){\line(1,0){90}}
\put(45,237){\line(0,-1){45}}
\put(135,192){\line(-1,0){90}}
\put(135,192){\line(0,1){45}}

\put(50,-12){Standard deviation of additive noise applied to training data}
\put(-10,70){\rotatebox{90}{Samples of training data}}

\end{picture}

\vspace{5mm}

\caption{In order to estimate the integral operator sufficiently well to achieve a given level of accuracy in held-out data, we require a different amount of training data depending on both the structure of matrix used to represent the operator as well as the level of noise present in the training data. We observe that, regardless of the level of noise present in the training data, the amount of experimental data required to achieve a given accuracy is roughly proportional to the number of parameters in the model, here following a roughly 1:8:16 ratio.}
\end{figure*}

This matrix $A$ can be approximated arbitrarily well by symmetric Toeplitz matrices~\citep{gray2006toeplitz}, hierarchical matrices, or dense matrices. For the sake of illustration, we choose to discretize the operator coarsely such that $A$ is a $16 \times 16$ matrix. With this level of discretization, a symmetric Toeplitz matrix takes 16 parameters to specify, a hierarchical matrix choosing $k=1$ takes 128, and a dense matrix takes 256. This is a ratio of 1:8:16 parameters for the three matrix types.

A common heuristic when building statistical models is that the amount of experimental data required to achieve a given generalization accuracy is roughly proportional to the number of parameters in the model. This can be motivated by considering the logarithm of Bayes' rule applied to the probability of a model given data, and interpreting it information theoretically. This argument is non-rigorous, but can be made rigorous under certain assumptions, for example in the case of parameter estimation for a multidimensional Gaussian distribution from random samples. We find that this heuristic holds surprisingly well in the case of identifying this discretized integral operator as can be seen in Fig.~\ref{fig:bayesian} and we will see that it continues to hold for a physical system identification problem in Sec.~\ref{sec:physical}. This means that model parameter reduction directly translates to a proportional reduction in the amount of experimental data required to achieve the same level of accuracy.

We now turn to an experimental demonstration of this form of statistical regularization used to practical effect for system identification.

\section{Physical experiment}
\label{sec:physical}
In this section we perform \textit{Coarse-Grained Nonlinear System Identification} on experimental data from a voltage~$V$ controlled tungsten filament, measuring luminosity~$L$ at a sampling rate of 750~Hz. Such a filament exhibits several nonlinearities. The resistance~$R$ of the filament is temperature~$T$ dependent, which affects the amount of voltage-driven heating. The filament cools primarily via convection and radiation which go as $T^2$ and $T^4$ respectively. This yields a state-space model in the single state variable $T$.
\begin{align*}
\dot T \: =& \: k_1 \frac{V^2}{R(T)} \: - \: k_2 T^2 \: - \: k_3 T^4\\
L \: =& \: k_4 T^4
\end{align*}
Other dynamics, for example driven by the inductance of the filament, are comparatively small at the timescale we consider.

\begin{figure}[!t]
\newcommand{\propx}{400}
\newcommand{\propy}{300}
\newcommand{\xaxx}{47}
\newcommand{\xaxy}{-6}
\newcommand{\yaxx}{-6}
\newcommand{\yaxy}{48}
\newcommand{\titley}{28}
\newcommand{\corrx}{119}
\newcommand{\corry}{70}
\newcommand{\corryaxx}{-5}
\newcommand{\corryaxy}{28}
\newcommand{\corrxaxx}{50}
\newcommand{\corrxaxy}{2}
\centering

\begin{picture}(\propx,310)
\put(0,0){\includegraphics[width=82mm]{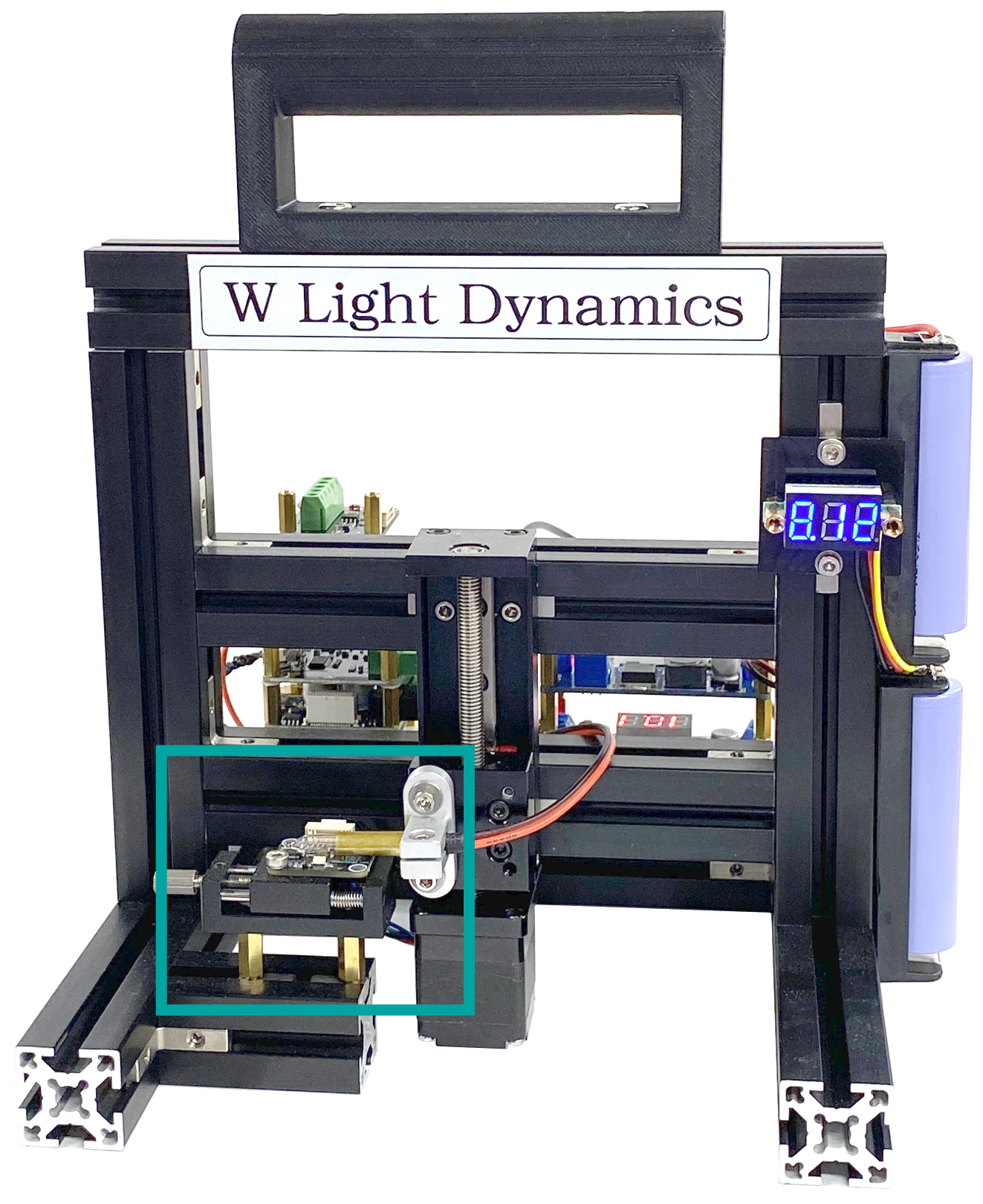}}
\put(53,300){\textbf{Experimental apparatus}}
\end{picture}
\begin{picture}(\propx,180)
\put(0,0){\includegraphics[width=82mm]{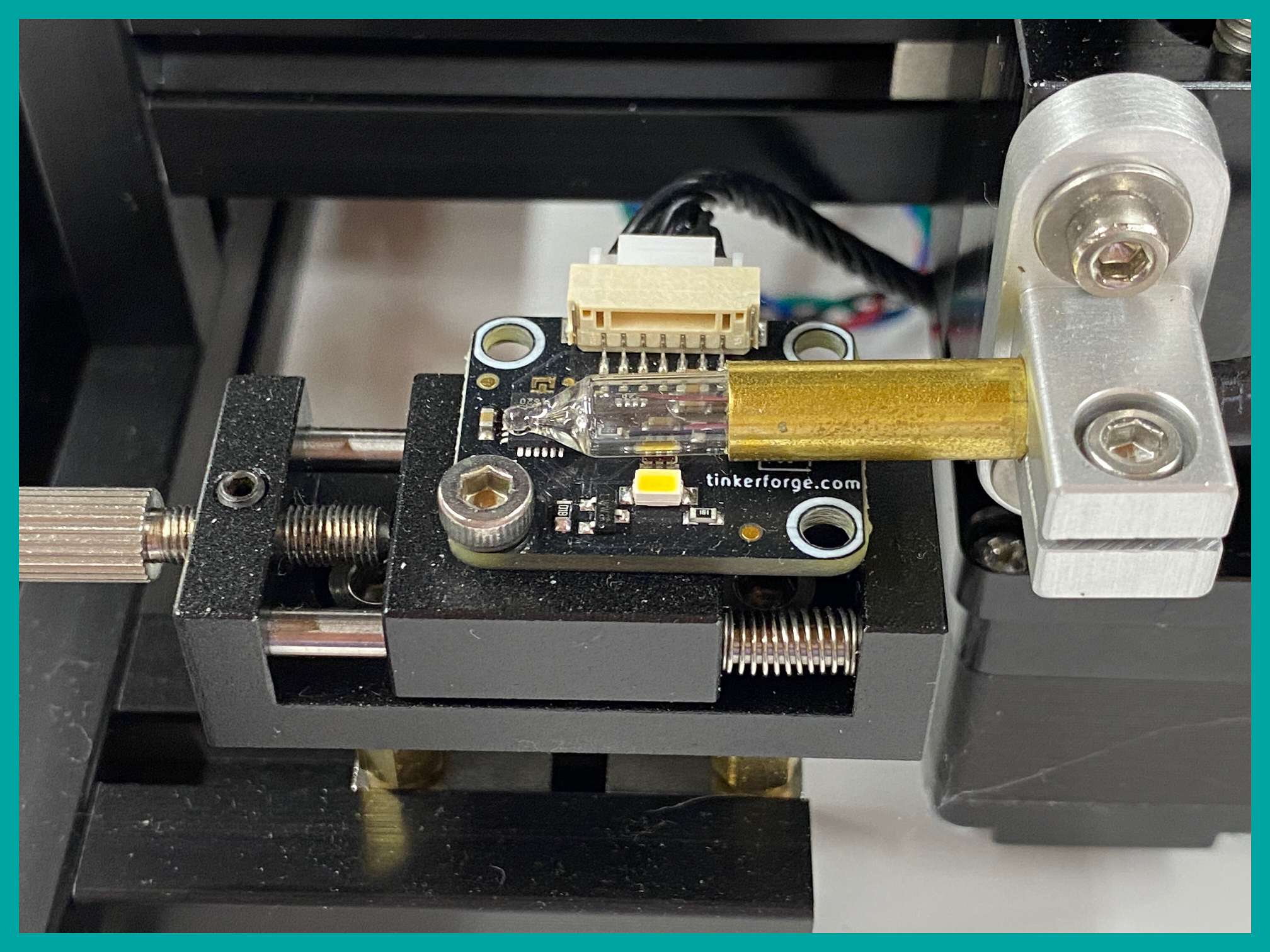}}
\end{picture}

\label{fig:mica}
\caption{MICA workstation for measuring the voltage-luminosity dynamics of a tungsten filament.}
\vspace{0cm}
\end{figure}

\subsection{Experimental apparatus}
We developed a MICA workstation~\citep{cheney2020development} to measure these dynamics, shown in Fig.~\ref{fig:mica}. MICA is a framework for rapidly building precision experimental systems primarily for use in education.

This system consists of a light sensor underneath a small diameter coiled tungsten filament light bulb. The light bulb is mounted on a computer controlled micro-stepping motor linear stage which allows the distance between the filament and the photosensor to be varied. In these experiments the filament to photosensor distance was set to  5 mm. A Digilent Discovery 2 was used to digitally record the output of the photosensor and was also used to output the signal sent to the tungsten filament from a digital-to-analog converter (14 bit DAC) via a linear power amplifier. The current (converted to a voltage) and voltage delivered through the tungsten filament together with the photosensor output current (converted to a voltage) were measured at 750 Hz for most of the experiments using simultaneous sampling analog-to-digital converters (variable gain, 14-bit ADCs). Both DAC and ADCs have very low inter-sample interval jitter ($<$5 ns).

\subsection{Input signal}
\begin{figure*}[!t]
\newcommand{\propx}{230}
\newcommand{\propy}{120}
\newcommand{\xaxx}{47}
\newcommand{\xaxy}{-6}
\newcommand{\yaxx}{-6}
\newcommand{\yaxy}{48}
\newcommand{\titley}{28}
\newcommand{\corrx}{119}
\newcommand{\corry}{70}
\newcommand{\corryaxx}{-5}
\newcommand{\corryaxy}{28}
\newcommand{\corrxaxx}{50}
\newcommand{\corrxaxy}{2}
\centering

\vspace{1mm}
\textbf{Model performance when trained on 1 s of experimental data}

\vspace{2mm}
\vspace{-2cm}
\hspace{2.5cm}
\begin{picture}(\propx,\propy)
\put(0,0){\includegraphics[width=80mm]{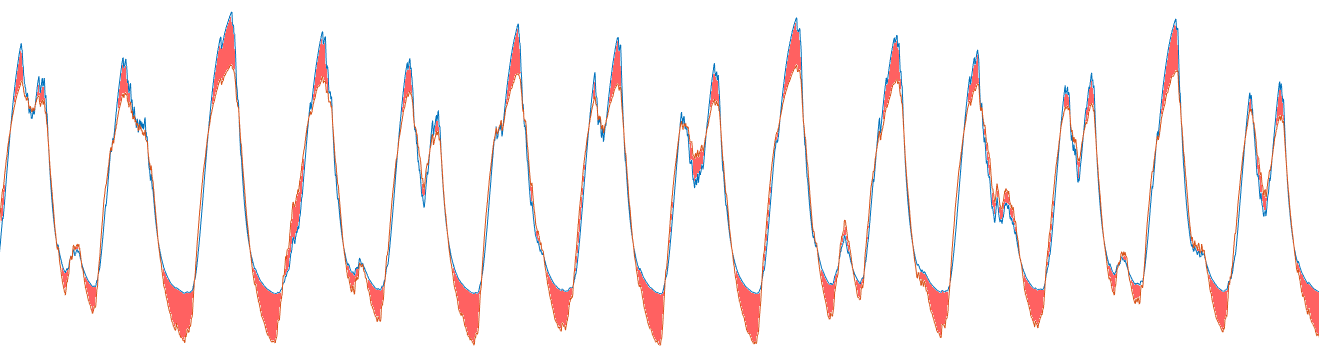}}
\put(-67,\titley){Linear model}
\end{picture}

\vspace{-2cm}
\hspace{2.5cm}
\begin{picture}(\propx,\propy)
\put(0,0){\includegraphics[width=80mm]{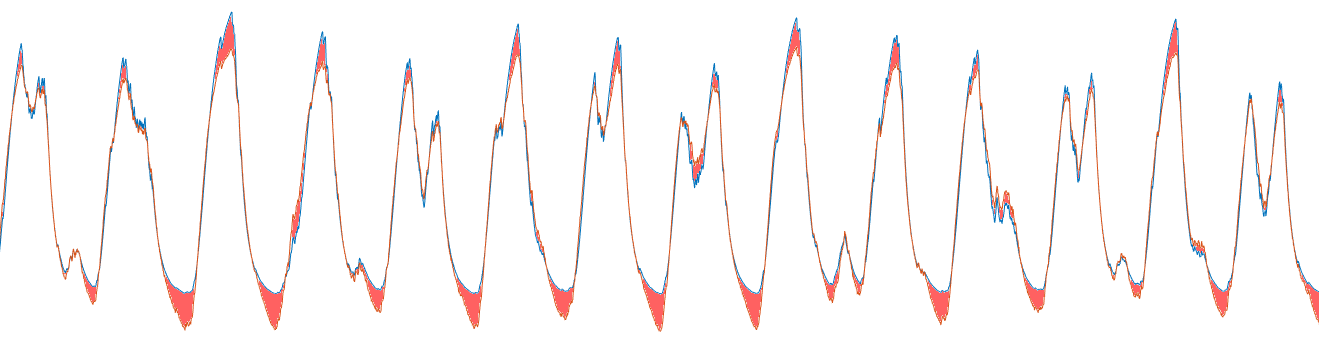}}
\put(-120,35.5){2\textsuperscript{nd} order Volterra model}
\put(-113.5,20.5){dense, $L_2$ regularization}
\end{picture}

\vspace{-2cm}
\hspace{2.5cm}
\begin{picture}(\propx,\propy)
\put(0,0){\includegraphics[width=80mm]{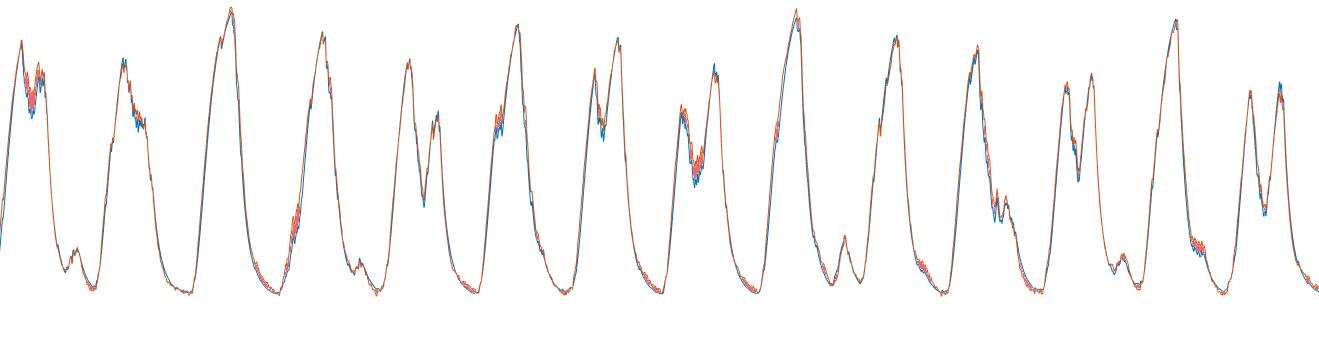}}
\put(-120,35.5){2\textsuperscript{nd} order Volterra model}
\put(-126,20.5){\textit{Coarse-Grained System ID}}
\end{picture}

\vspace{-3.5cm}
\hspace{2.5cm}
\begin{picture}(\propx,\propy)
\put(0,0){\includegraphics[width=80mm]{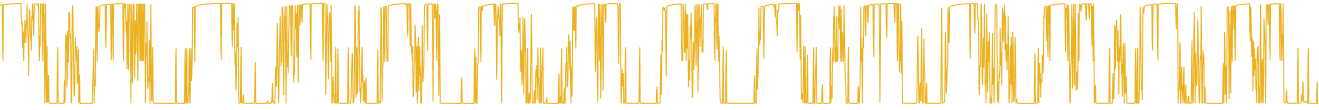}}
\put(-62,8){Input signal}
\end{picture}

\vspace{-3.8cm}
\hspace{2.5cm}
\begin{picture}(\propx,\propy)
\put(0,0){\includegraphics[width=80mm]{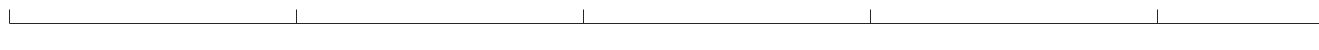}}
\put(90,-19){Time (s)}

\put(-1,-5){0}
\put(49,-5){1}
\put(98.5,-5){2}
\put(148,-5){3}
\put(197.5,-5){4}

\end{picture}

\vspace{7mm}

\caption{Performance of various model types on a held-out data set when trained on only 1~s of data at 750 Hz. The blue line is experimental data, the orange line is the model's prediction, and the red regions highlight the error. The linear model cannot capture the nonlinearities in the system dynamics and achieves only 93.5\% variance accounted for. The dense model would be badly overfit if not for extreme levels of $L_2$ regularization; at an optimally tuned level of $L_2$ regularization it achieves 97.5\% variance accounted for. \textit{Coarse-Grained Nonlinear System ID} is a much more effective form of regularization, achieving 99.4\% variance accounted for.}
\label{fig:exp-results-time}
\end{figure*}
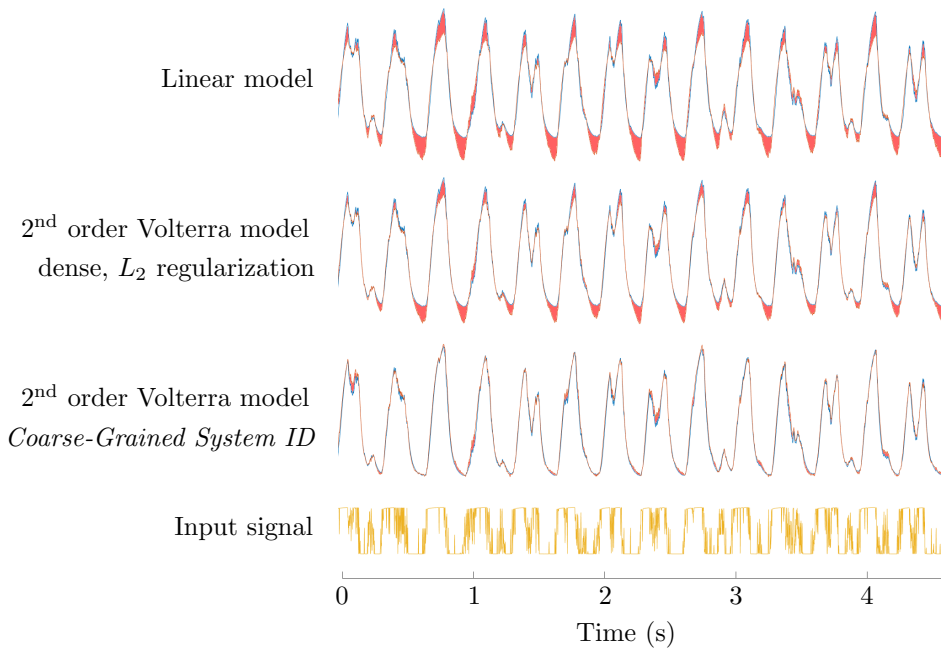

In order to excite the tungsten filament through the full range of its nonlinear dynamics, we chose an input voltage signal with a tuned power spectral density while satisfying constraints on the range of possible input values. This signal was generated using a novel combination of optimization and stochastic interchange described by~\citet{spanbauer2020rapid} in a companion paper. A characteristic portion of the input signal can be seen in Fig.~\ref{fig:exp-results-time}

\subsection{Model details}
We used a second order Volterra model with a system memory of 128~samples. At the sampling rate of 750~Hz this corresponds to a memory of about 0.17~s, which we found to be sufficient. 

For the hierarchical matrix representation of the second order Volterra kernel, we chose the parameter controlling the number of vector outer products involved in the low-rank approximations, $k$, to be 1. This leads to a model parameter reduction of about 8.5$\times$, from 16384 for the dense model to 1920 for the hierarchical matrix model.

In all experiments we tuned a single hyperparameter, the level of $L_2$ regularization, to achieve the best possible performance on a held-out test set. Very strong $L_2$ regularization was required to achieve reasonable performance when using a dense representation of the second order Volterra kernel even when relatively large amounts (30~s) of training data was available. When using \textit{Coarse-Grained Nonlinear System ID}, mild $L_2$ regularization was helpful when only small amounts of training data (1~s) were available; when larger amounts of training data were available no $L_2$ regularization was required.

Models were trained to convergence in PyTorch~\citep{NEURIPS2019_9015} using an Adam optimizer~\citep{kingma2014adam}, which typically took less than two minutes on an NVIDIA~RTX~2080~Ti GPU.

\subsection{Results}

\textit{Coarse-Grained Nonlinear System Identification} achieves a very strong regularization effect consistent with its reduction in the number of model parameters by almost an order of magnitude without loss in representational power.

We observe that our method learns the nonlinear voltage to luminosity dynamics of the tungsten filament accurately with only one second of data, unlike the traditionally regularized dense model as can be seen in Fig.~\ref{fig:exp-results-time}.

\begin{figure*}[!t]
\newcommand{\propx}{400}
\newcommand{\propy}{300}
\newcommand{\xaxx}{47}
\newcommand{\xaxy}{-6}
\newcommand{\yaxx}{-6}
\newcommand{\yaxy}{48}
\newcommand{\titley}{28}
\newcommand{\corrx}{119}
\newcommand{\corry}{70}
\newcommand{\corryaxx}{-5}
\newcommand{\corryaxy}{28}
\newcommand{\corrxaxx}{50}
\newcommand{\corrxaxy}{2}
\centering

\begin{picture}(\propx,240)
\put(0,0){\includegraphics[width=155mm]{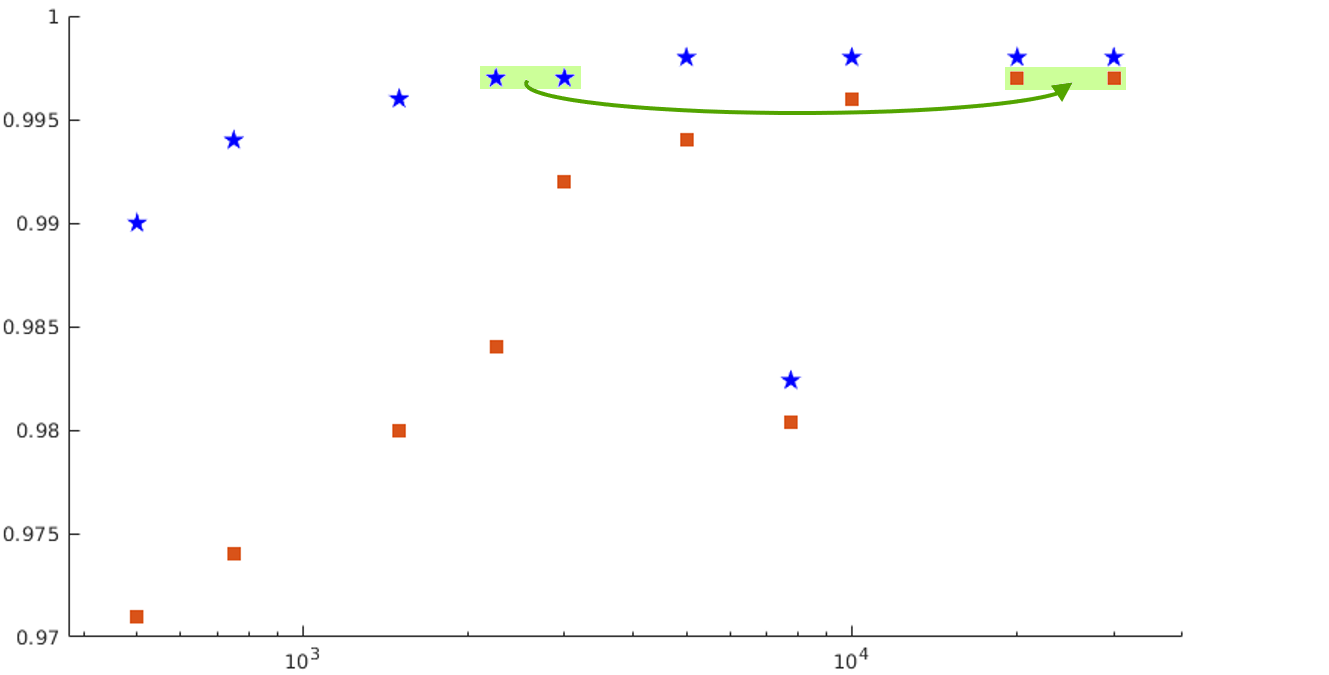}}
\put(65,230){\textbf{Regularization effect for varying amounts of training data}}
\put(273,94.5){Coarse-grained system ID}
\put(273,80.5){Dense, $L_2$ regularization}
\put(254,75){\line(1,0){138}}
\put(254,107){\line(1,0){138}}
\put(254,107){\line(0,-1){32}}
\put(392,107){\line(0,-1){32}}
\put(-20,20){\rotatebox{90}{Variance accounted for in held-out data set}}
\put(120,-10){Samples of training data at 750 Hz}
\definecolor{chart_green}{RGB}{82, 164, 0}
\put(240,175){\textcolor{chart_green}{\textbf{1/10\textsuperscript{th} the data for the same accuracy}}}
\end{picture}

\vspace{5mm}

\caption{\textit{Coarse-Grained Nonlinear System ID} outperforms $L_2$ regularization at all amounts of training data. In the limit of large amounts of training data, both methods converge to a variance accounted for of 99.8\%; the remaining variance is either noise or higher-order nonlinear dynamics. We observe rough agreement with the heuristic described in Sec.~\ref{sec:synthetic}: our hierarchical matrix model has roughly 10x fewer parameters than our dense model, and we observe that roughly 10x less data is required to achieve the same generalization accuracy.}
\label{fig:exp-results-reg}
\end{figure*}
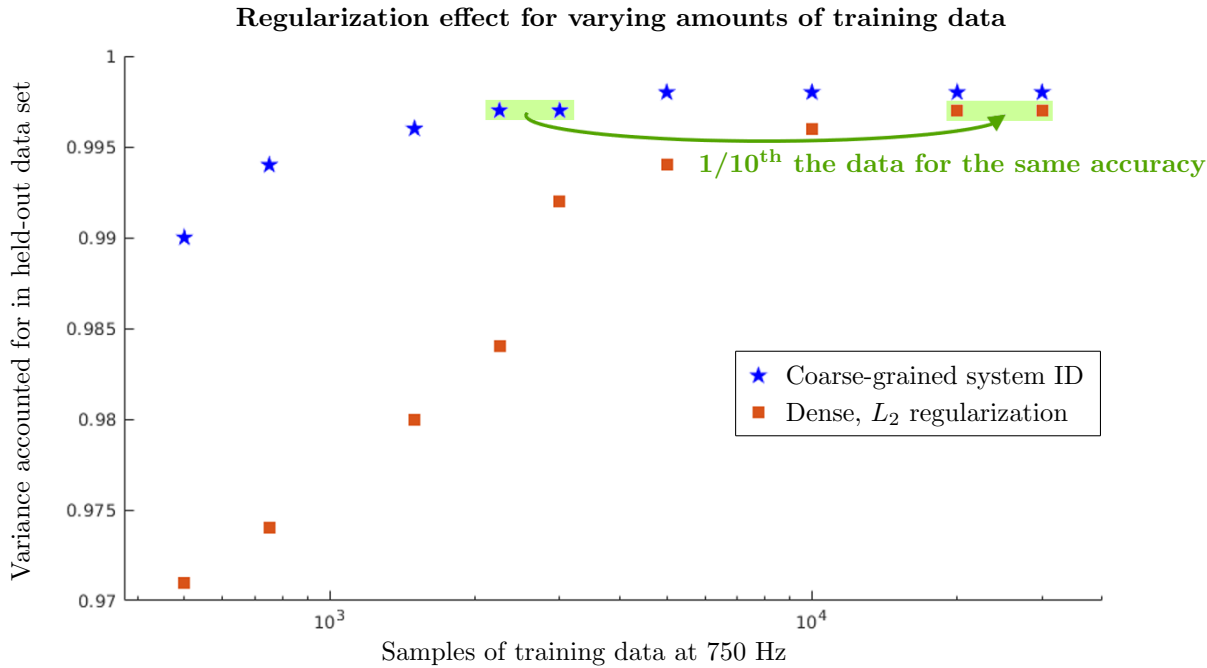

This improvement over the traditionally regularized dense model holds for any amount of training data as can be seen in Fig.~\ref{fig:exp-results-reg}. We find that for this system \textit{Coarse-Grained Nonlinear System Identification} requires almost an order of magnitude less training data than a traditionally regularized Volterra model to achieve the same level of generalization accuracy.

\begin{figure}[!t]
\newcommand{\propx}{400}
\newcommand{\propy}{300}
\newcommand{\xaxx}{47}
\newcommand{\xaxy}{-6}
\newcommand{\yaxx}{-6}
\newcommand{\yaxy}{48}
\newcommand{\titley}{28}
\newcommand{\corrx}{119}
\newcommand{\corry}{70}
\newcommand{\corryaxx}{-5}
\newcommand{\corryaxy}{28}
\newcommand{\corrxaxx}{50}
\newcommand{\corrxaxy}{2}
\centering

\vspace{0.5cm}
\begin{picture}(196,186)
\put(0,-20){\includegraphics[width=70mm]{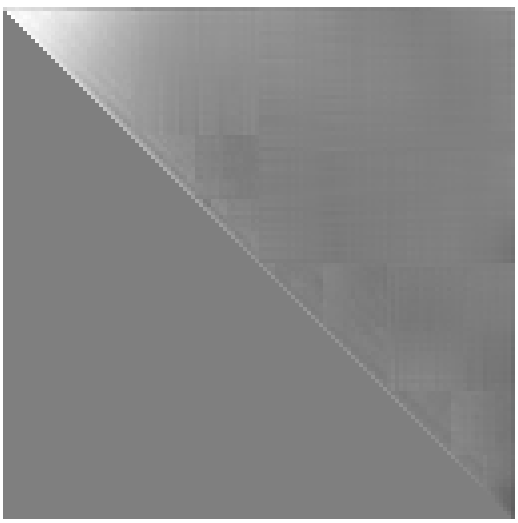}}
\put(7,190){\textbf{Identified 2\textsuperscript{nd} order Volterra Kernel}}
\end{picture}

\vspace{5mm}
\caption{This linearly scaled heatmap depicts a second order Volterra kernel estimated via \textit{Coarse-Grained System Identification} using 30~s of training data. There is no ground-truth second order Volterra kernel in the case of no $L_2$ regularization, since Volterra kernels do not form an orthogonal basis; this is an optimized kernel in the presence of mild $L_2$ regularization.}
\vspace{-8mm}
\label{fig:kernel}
\end{figure}

The identified second order Volterra kernel for this system is shown in Fig.~\ref{fig:kernel}.

\section{Discussion}

This paper has introduced \textit{Coarse-Grained Nonlinear Dynamics}, a new method of parameterizing nonlinear dynamic models achieving a superpolynomial reduction in the number of parameters required to specify the model compared to the traditional dense representation of Volterra kernels. This represents a strict reduction in the scaling of the number of parameters from the multiparameter complexity class \textbf{XP} to the class \textbf{FPT}.

\vspace{0.4mm}
This superpolynomial reduction in number of model parameters is leveraged as a statistical regularizer in \textit{Coarse-Grained Nonlinear System Identification}, directly translating to a proportional decrease in the amount of experimental data required to achieve a certain level of model accuracy. This property has held up in the analysis of experimental data, allowing us to accurately learn the nonlinear voltage to luminosity dynamics of a tungsten filament with only one second of data.

\vspace{0.4mm}
These results create the potential for fast learning of third- and higher-order Volterra kernels, a largely unexplored space due the the rapid exponential scaling in experimental data required to estimate these kernels using past methods. Despite difficulties in estimation, higher order kernels have seen use. For example \citet{wong1990discrete,wong1990discreteb} showed that all even-order Volterra kernels are zero for a nuclear magnetic resonance (NMR) spin system and went on to analyze the NMR physics as it contributes to the third-order kernel. They determined first- through third-order Volterra kernels using first- and higher-order cross-correlation estimates from data collected using a stochastically excited probe coil in a 0.5 T magnetic field.

Another example of an application area is in the field of artificial muscle fiber actuators. A number of these actuators including fast contracting shape memory alloy actuators~\citep{hunter1992shape} and contractile conducting polymers~\citep{madden2004artificial,mirvakili2018artificial} display highly nonlinear dynamic behavior which has been a challenge to model (and then control) using traditional experimental data based modeling approaches.

Considering the success that Volterra kernels have had throughout the sciences, the authors are optimistic that \textit{Coarse-Grained Nonlinear System Identification} will provide fast system identification for a wide-range of physical phenomena, including difficult-to-model phenomena exhibiting high-order nonlinearities.

\section*{Acknowledgements}

The authors would like to thank Carlos Perez-Arancibia for helpful instruction and discussions, and Fonterra for supporting this research.


\newpage

\bibliography{mybib}

\begin{thebibliography}{32}
\providecommand{\natexlab}[1]{#1}
\providecommand{\url}[1]{\texttt{#1}}
\expandafter\ifx\csname urlstyle\endcsname\relax
  \providecommand{\doi}[1]{doi: #1}\else
  \providecommand{\doi}{doi: \begingroup \urlstyle{rm}\Url}\fi

\bibitem[Asfour et~al.(2000)Asfour, Raoof, and Fournier]{asfour2000nonlinear}
A.~Asfour, K.~Raoof, and J.-M. Fournier.
\newblock {Nonlinear identification of NMR spin systems by adaptive filtering}.
\newblock \emph{Journal of Magnetic Resonance}, 145\penalty0 (1):\penalty0
  37--51, 2000.

\bibitem[Biswas and McGee(1991)]{biswas1991volterra}
T.~K. Biswas and W.~McGee.
\newblock {Volterra series analysis of semiconductor laser diode}.
\newblock \emph{IEEE Photonics Technology Letters}, 3\penalty0 (8):\penalty0
  706--708, 1991.

\bibitem[Chen and Hunter(2013)]{chen2013nonlinear}
Y.~Chen and I.~W. Hunter.
\newblock {Nonlinear stochastic system identification of skin using Volterra
  kernels}.
\newblock \emph{Annals of biomedical engineering}, 41\penalty0 (4):\penalty0
  847--862, 2013.

\bibitem[Cheney(2020)]{cheney2020development}
C.~Cheney.
\newblock \emph{{Development of a miniature, low power, solid state,
  continuously sensitive, diffusion cloud chamber}}.
\newblock PhD thesis, Massachusetts Institute of Technology, 2020.

\bibitem[Downey and Fellows(2013)]{downey2013fundamentals}
R.~G. Downey and M.~R. Fellows.
\newblock \emph{{Fundamentals of parameterized complexity}}, volume~4.
\newblock Springer, 2013.

\bibitem[Gray(2006)]{gray2006toeplitz}
R.~M. Gray.
\newblock \emph{{Toeplitz and circulant matrices: A review}}.
\newblock now publishers inc, 2006.

\bibitem[Gruber et~al.(2012)Gruber, Bordons, and Oliva]{gruber2012nonlinear}
J.~Gruber, C.~Bordons, and A.~Oliva.
\newblock {Nonlinear MPC for the airflow in a PEM fuel cell using a Volterra
  series model}.
\newblock \emph{Control Engineering Practice}, 20\penalty0 (2):\penalty0
  205--217, 2012.

\bibitem[Hackbusch(1990)]{hackbusch1990panel}
W.~Hackbusch.
\newblock {The panel clustering algorithm}.
\newblock In \emph{MAFELAP}, pages 339--348, 1990.

\bibitem[Hackbusch(1999)]{hackbusch1999}
W.~Hackbusch.
\newblock {A sparse matrix arithmetic based on $\cal h$-matrices. part i:
  Introduction to $\cal h$-matrices}.
\newblock \emph{Computing}, 62\penalty0 (2):\penalty0 89--108, 1999.

\bibitem[Hunter and Lafontaine(1992)]{hunter1992shape}
I.~Hunter and S.~R. Lafontaine.
\newblock {Shape memory alloy fibers having rapid twitch response}, Mar.~3
  1992.
\newblock US Patent 5,092,901.

\bibitem[Hunter and Lafontaine(2016)]{hunter2016apparatus}
I.~Hunter and S.~R. Lafontaine.
\newblock {Apparatus and method for rapidly charging batteries}, July~19 2016.
\newblock US Patent 9,397,516.

\bibitem[Joeken et~al.(1997)Joeken, Schwegler, and Richter]{joeken1997modeling}
S.~Joeken, H.~Schwegler, and C.-P. Richter.
\newblock {Modeling stochastic spike train responses of neurons: an extended
  Wiener series analysis of pigeon auditory nerve fibers}.
\newblock \emph{Biological Cybernetics}, 76\penalty0 (2):\penalty0 153--162,
  1997.

\bibitem[Kaizer(1987)]{kaizer1987modeling}
A.~J. Kaizer.
\newblock {Modeling of the nonlinear response of an electrodynamic loudspeaker
  by a Volterra series expansion}.
\newblock \emph{Journal of the Audio Engineering Society}, 35\penalty0
  (6):\penalty0 421--433, 1987.

\bibitem[Kamalakis et~al.(2011)Kamalakis, Walewski, Ntogari, and
  Mileounis]{kamalakis2011empirical}
T.~Kamalakis, J.~W. Walewski, G.~Ntogari, and G.~Mileounis.
\newblock {Empirical Volterra-series modeling of commercial light-emitting
  diodes}.
\newblock \emph{Journal of Lightwave Technology}, 29\penalty0 (14):\penalty0
  2146--2155, 2011.

\bibitem[Kingma and Ba(2014)]{kingma2014adam}
D.~P. Kingma and J.~Ba.
\newblock {Adam: A method for stochastic optimization}.
\newblock \emph{arXiv preprint arXiv:1412.6980}, 2014.

\bibitem[Korenberg and Hunter(1990)]{korenberg1990identification}
M.~J. Korenberg and I.~W. Hunter.
\newblock {The identification of nonlinear biological systems: Wiener kernel
  approaches}.
\newblock \emph{Annals of Biomedical Engineering}, 18\penalty0 (6):\penalty0
  629--654, 1990.

\bibitem[Korenberg and Hunter(1996)]{korenberg1996identification}
M.~J. Korenberg and I.~W. Hunter.
\newblock {The identification of nonlinear biological systems: Volterra kernel
  approaches}.
\newblock \emph{Annals of biomedical engineering}, 24\penalty0 (2):\penalty0
  250--268, 1996.

\bibitem[Lafontaine and Hunter(2018)]{lafontaine2018nonlinear}
S.~R. Lafontaine and I.~W. Hunter.
\newblock {Nonlinear system identification for object detection in a wireless
  power transfer system}, May~29 2018.
\newblock US Patent 9,983,243.

\bibitem[Litt(1997)]{litt1997nonlinear}
H.~I. Litt.
\newblock \emph{{A nonlinear kernel investigation of magnetic resonance imaging
  and computed tomography}}.
\newblock PhD thesis, 1997.

\bibitem[Madden et~al.(2004)Madden, Vandesteeg, Anquetil, Madden, Takshi,
  Pytel, Lafontaine, Wieringa, and Hunter]{madden2004artificial}
J.~D. Madden, N.~A. Vandesteeg, P.~A. Anquetil, P.~G. Madden, A.~Takshi, R.~Z.
  Pytel, S.~R. Lafontaine, P.~A. Wieringa, and I.~W. Hunter.
\newblock {Artificial muscle technology: physical principles and naval
  prospects}.
\newblock \emph{IEEE Journal of oceanic engineering}, 29\penalty0 (3):\penalty0
  706--728, 2004.

\bibitem[Marmarelis and Naka(1973)]{marmarelis1973nonlinear}
P.~Z. Marmarelis and K.~Naka.
\newblock {Nonlinear analysis and synthesis of receptive-field responses in the
  catfish retina. I. Horizontal cell leads to ganglion cell chain}.
\newblock \emph{Journal of Neurophysiology}, 36\penalty0 (4):\penalty0
  605--618, 1973.

\bibitem[Mirvakili and Hunter(2018)]{mirvakili2018artificial}
S.~M. Mirvakili and I.~W. Hunter.
\newblock {Artificial muscles: Mechanisms, applications, and challenges}.
\newblock \emph{Advanced Materials}, 30\penalty0 (6):\penalty0 1704407, 2018.

\bibitem[Narayanan(1970)]{narayanan1970application}
S.~Narayanan.
\newblock {Application of Volterra series to intermodulation distortion
  analysis of transistor feedback amplifiers}.
\newblock \emph{IEEE Transactions on Circuit Theory}, 17\penalty0 (4):\penalty0
  518--527, 1970.

\bibitem[Paszke et~al.(2019)Paszke, Gross, Massa, Lerer, Bradbury, Chanan,
  Killeen, Lin, Gimelshein, Antiga, Desmaison, Kopf, Yang, DeVito, Raison,
  Tejani, Chilamkurthy, Steiner, Fang, Bai, and Chintala]{NEURIPS2019_9015}
A.~Paszke, S.~Gross, F.~Massa, A.~Lerer, J.~Bradbury, G.~Chanan, T.~Killeen,
  Z.~Lin, N.~Gimelshein, L.~Antiga, A.~Desmaison, A.~Kopf, E.~Yang, Z.~DeVito,
  M.~Raison, A.~Tejani, S.~Chilamkurthy, B.~Steiner, L.~Fang, J.~Bai, and
  S.~Chintala.
\newblock {PyTorch: An Imperative Style, High-Performance Deep Learning
  Library}.
\newblock In H.~Wallach, H.~Larochelle, A.~Beygelzimer, F.~d\textquotesingle
  Alch\'{e}-Buc, E.~Fox, and R.~Garnett, editors, \emph{Advances in Neural
  Information Processing Systems 32}, pages 8024--8035. Curran Associates,
  Inc., 2019.

\bibitem[Rokhlin(1985)]{rokhlin1985}
V.~Rokhlin.
\newblock {Rapid solution of integral equations of classical potential theory}.
\newblock \emph{Journal of computational physics}, 60\penalty0 (2):\penalty0
  187--207, 1985.

\bibitem[Sandberg and Stark(1968)]{sandberg1968wiener}
A.~Sandberg and L.~Stark.
\newblock {Wiener G-function analysis as an approach to non-linear
  characteristics of human pupil light reflex}.
\newblock \emph{Brain Research}, 11\penalty0 (1):\penalty0 194--211, 1968.

\bibitem[Spanbauer and Hunter(2020)]{spanbauer2020rapid}
S.~Spanbauer and I.~Hunter.
\newblock {Rapid Generation of Stochastic Signals with Specified Statistics}.
\newblock \emph{arXiv preprint}, 2020.

\bibitem[Volterra(1887)]{volterra1887}
V.~Volterra.
\newblock \emph{{Sopra le funzioni che dipendono da altre funzioni}}.
\newblock Tip. della R. Accademia dei Lincei, 1887.
\newblock URL \url{https://books.google.com/books?id=8zKnAQAACAAJ}.

\bibitem[Wiener(1966)]{wiener1966nonlinear}
N.~Wiener.
\newblock \emph{{Nonlinear problems in random theory}}.
\newblock 1966.

\bibitem[Wong et~al.(1990{\natexlab{a}})Wong, Rods, Newmark, and
  Budinger]{wong1990discreteb}
S.~Wong, M.~Rods, R.~Newmark, and T.~Budinger.
\newblock {Discrete analysis of stochastic NMR. II}.
\newblock \emph{Journal of Magnetic Resonance (1969)}, 87\penalty0
  (2):\penalty0 265--286, 1990{\natexlab{a}}.

\bibitem[Wong et~al.(1990{\natexlab{b}})Wong, Rods, Newmark, and
  Budinger]{wong1990discrete}
S.~T.-S. Wong, M.~Rods, R.~Newmark, and T.~Budinger.
\newblock {Discrete analysis of stochastic NMR. I}.
\newblock \emph{Journal of Magnetic Resonance (1969)}, 87\penalty0
  (2):\penalty0 242--264, 1990{\natexlab{b}}.

\bibitem[Zhu and Brazil(2004)]{zhu2004behavioral}
A.~Zhu and T.~J. Brazil.
\newblock {Behavioral modeling of RF power amplifiers based on pruned Volterra
  series}.
\newblock \emph{IEEE Microwave and Wireless components letters}, 14\penalty0
  (12):\penalty0 563--565, 2004.

\end{thebibliography}

\end{document}